\documentclass{ws-procs9x6}

\begin{document}


\newcommand{\CT}{S} 
\newcommand\ZZ{\mathcal{Z}}
\newcommand\CurlD{\mathcal{D}}
\newcommand\MM{\mathcal{M}}
\newcommand\dM{\partial \MM}
\newcommand\bns{\!\!\!}

\title{PATH INTEGRAL FOR HALF-BINDING POTENTIALS \\ AS QUANTUM MECHANICAL ANALOG FOR \\ BLACK HOLE PARTITION FUNCTIONS}

\author{D. GRUMILLER}

\address{Center for Theoretical Physics, Massachusetts Institute of Technology,\\
77 Massachusetts Ave., Cambridge, MA 02139, USA \\
E-mail: grumil@lns.mit.edu}

\begin{abstract}
The semi-classical approximation to black hole partition functions is not well-defined, because the classical action is unbounded and the first variation of the uncorrected action does not vanish for all variations preserving the boundary conditions. Both problems can be solved by adding a Hamilton-Jacobi counter\-term. 
I show that the same problem and solution arises in quantum mechanics for half-binding potentials.
\end{abstract}


\bodymatter

\section{Introduction and statement of the problem}\label{se:1}

Path integrals have illuminated many aspects of quantum mechanics and quantum field theory \cite{Dresden:2007}, but there remain some challenges to path integral formulations of quantum theories \cite{Jackiw:2007bc}. In this proceedings contribution I describe a problem arising for quantum mechanical potentials that are `half-binding' (the definition of this term will be given below). I shall demonstrate that the naive semi-classical approximation to the path integral breaks down for two reasons: the leading contribution to the partition function is singular and the first variation of the action does not vanish for all variations preserving the boundary conditions. I discuss how both issues can be resolved by adding an appropriate (Hamilton-Jacobi) counterterm as boundary term to the action. Moreover, I shall point out formal similarities to black hole (BH) partition functions, so in that sense these quantum mechanical systems may serve as toy models to elucidate certain aspects of BH physics.
For sake of clarity I focus on a specific Hamiltonian \cite{deAlfaro:1976je}, 
\begin{equation}
  \label{eq:dresden1}
  H(q,p)=\frac{p^2}{2}+V(q)\,,\qquad V(q)=\frac{1}{q^2}\,,
\end{equation}
where $q$ is restricted to positive values. If $q$ is small the Hamiltonian rises without bound, like for a binding potential. If $q$ is large the potential is negligible, and the asymptotic dynamics is dominated by free propagation. I refer to a potential with these properties as `half-binding'. [The conformal properties \cite{deAlfaro:1976je} of \eqref{eq:dresden1} will not play any role in this discussion.]

Consistency of the variational principle based on the Lagrangian action,
\begin{equation}
  \label{eq:dresden2}
  I[q]=\int\limits_{t_i}^{t_f}\!dt \,\Big(\frac{\dot{q}^2}{2}-\frac{1}{q^2}\Big) \,,
\end{equation}
requires to fix the initial and final value of $q$ at $t_i$ and $t_f$, respectively. I am interested here mostly in the limit $t_f\to\infty$, which implies that $q|^{t_f}=\infty$ is the appropriate asymptotic boundary condition. The initial time is set to zero, $t_i=0$, without loss of generality.
The Lagrangian path integral,
\begin{equation}
  \label{eq:dresden3}
   \ZZ = \int \CurlD q \, \exp\Big(-\frac{1}{\hbar}\,I[q] \Big)\,, 
\end{equation}
consists of a coherent sum over all field configurations consistent with the boundary data. Even though \eqref{eq:dresden2} is exactly soluble, it is illustrative to consider the semi-classical expansion of the action,
\begin{equation}
  \label{eq:dresden4}
  I[q_{\rm cl} + \delta q] = I[q_{\rm cl}] + \delta I\big|_{\rm EOM}  + {\cal O}(\delta q^2) \,,
\end{equation}
and of the partition function
\begin{equation}
  \label{eq:dresden5}
  \ZZ = \exp\Big(-\frac{1}{\hbar}\,I[q_{\rm cl}]\Big) \, \int \CurlD \delta q \, \exp\Big(-\frac{1}{\hbar}\,{\cal O}(\delta q^2) \Big) \,.
\end{equation}
The semi-classical approximation \eqref{eq:dresden5} is well-defined only if the on-shell action is bounded, $|I[q_{\rm cl}]|<\infty$, and only if the first variation of the action vanishes on-shell, $\delta I|_{\rm EOM} = 0$, for all field configurations preserving the boundary conditions. I demonstrate now that neither is the case for the example \eqref{eq:dresden2}. 

The on-shell action diverges because asymptotically the propagation is essentially free, and because of the assumption $t_f\to\infty$. This is an idealization of situations where boundary conditions are imposed at late times, $t_f\sim 1/\epsilon$, with $\epsilon\ll 1$. In that case also $q_f\sim 1/\epsilon$ classically. However, the path integral does not only take into account classical contributions, but also samples nearby field configurations whose asymptotic behavior is $q \sim q_f [1 + \epsilon \, \Delta q + {\cal O}(\epsilon^2)]$, where $\Delta q$ is finite.
Therefore, the first variation of the action, evaluated on-shell, is given by the boundary term
\begin{equation}
  \label{eq:dresden5.5}
  \dot{q}\,\delta q|^{t_f} - \dot{q}\,\delta q|^{t_i=0} = \dot{q}\,\delta q|^{t_f} \sim \lim_{\epsilon\to 0} [\dot{q}\, \Delta q + {\cal O}(\epsilon)]\big|^{t_f}\neq 0\,.
\end{equation}
The inequality emerges, because 
arbitrary finite variations $\delta q|^{t_f}$ certainly preserve the boundary condition $q|^{t_f}=\infty$.\footnote{Canonical transformations can shift the problem, but of course they cannot solve it. For instance, with $Q=1/q$ and $P=-pq^2$ the correct asymptotic boundary condition is $Q|^{t_f}=0$ and therefore also $\delta Q|^{t_f}=0$. But now the momentum $P$ (and thus $\dot{Q}$) diverges at the boundary, so that the expression $\dot{Q}\,\delta Q|^{t_f}$ becomes undefined and does not necessarily vanish for all variations preserving the boundary conditions.} The two problems described here spoil the semi-classical approximation \eqref{eq:dresden5} to the partition function.

\section{Hamilton-Jacobi counterterm for half-binding potentials}\label{se:2}

Both problems can be solved by considering an improved action\footnote{There exists a variety of subtraction methods in quantum mechanics \cite{Kleinert}, in General Relativity (and generalizations thereof) \cite{Regge:1974zd} 
and in holographic renormalization within the context of AdS/CFT \cite{Henningson:1998gx}. 
Many of them have ad-hoc elements and require the subtraction of the action evaluated on a specific field configuration (like the ground state solution); in some cases there are several ``natural'' candidates, in others there is none, and in at least one example the ``natural'' guess even turned out to be wrong \cite{Davis:2004xiNote}.}
\begin{equation}
  \label{eq:dresden6}
  \Gamma[q] =\int\limits_{0}^{t_f}\! dt \,\Big(\frac{\dot{q}^2}{2}-\frac{1}{q^2}\Big) - \CT(q,t)\Big|^{t_f}_{0}\,,
\end{equation}
which differs from \eqref{eq:dresden2} by a boundary counterterm depending solely on quantities that are kept fixed at the boundary. The variation of \eqref{eq:dresden6},
\begin{equation}
  \label{eq:dresden7}
  \delta \Gamma|_{\rm EOM} = \Big(\dot{q}-\frac{\partial \CT}{\partial q}\Big)\delta q\Big|_{0}^{t_f} = \Big(\dot{q}-\frac{\partial \CT}{\partial q}\Big)\delta q\Big|^{t_f}\,,
\end{equation}
does not necessarily suffer from the second problem if $\partial \CT/\partial q$ asymptotically behaves like $\dot{q}$, i.e., like the momentum $p$.

The method \cite{deBoer:1999xf, Grumiller:2007ju} 
that I am going to review does not involve the subtraction of the action evaluated on a specific field configuration, but rather is intrinsic. Moreover, the amount of guesswork is minimal: Hamilton's principal function is a well-known function of the boundary data with the property $\partial\CT/\partial q= p$. Therefore it is natural to postulate that $\CT$  in \eqref{eq:dresden6} solves the Hamilton-Jacobi equation,
\begin{equation}
  \label{eq:dresden8}
  H\Big(q,\frac{\partial \CT}{\partial q}\Big)+\frac{\partial \CT}{\partial t} = 0\,.
\end{equation}

The complete integral \cite{Kamke}
\begin{equation}
  \label{eq:dresden9}
  \CT(q,t)= c_0 -Et+\sqrt{2(Eq^2-1)}+\sqrt{2}\,\arctan{\frac{1}{\sqrt{Eq^2-1}}}
\end{equation}
allows to construct the enveloping solution\footnote{\label{fn:1} One is forced to take the enveloping solution, since $\CT$ is part of the definition of the improved action and therefore cannot depend on constants of motion. The energy $E$ is eliminated from \eqref{eq:dresden9} by solving $\partial\CT/\partial E=0$ for $E$. The other constant, $c_0$, is set to zero by hand, but other choices are possible. Such an ambiguity always remains in this (and any other) approach. It reflects the freedom to shift the free energy of the ground state.}
\begin{equation}
  \label{eq:dresden10}
  \CT(q,t)=\frac{q^2}{2t} \left(\sqrt{4\Delta_+-8t^2/q^4}-\Delta_+\right)+\sqrt{2}\,\arctan{\frac{1}{\sqrt{q^4\Delta_+/(2t^2)-1}}}\,,
\end{equation}
where $\Delta_+ := \frac12 (1+\sqrt{1-8t^2/q^4})$.
The asymptotic expansion $\CT=q^2/(2t)+{\cal O}(t/q^2)$ is consistent with the intuitive idea that the asymptotically free propagation is the source of all subtleties. But the expression \eqref{eq:dresden10} contains a great deal of additional (non-asymptotic) information, which can be physically relevant, as mentioned in the next Section. 

Let me now come back to the two problems. Since asymptotically $\dot{q}|_{\rm EOM}=v=\rm const.$, the on-shell action
\begin{equation}
  \label{eq:dresden11}
  \Gamma\big|_{\rm EOM} = \frac{v^2}{2}\int\limits^{t_f}_0 \!dt - \frac{v^2}{2} t^f + {\cal O}(1) = {\cal O}(1) 
\end{equation}
evidently is finite. The terms of order of unity entail the information about the potential $V(q)$. The first variation
\begin{equation}
  \label{eq:dresden12}
  \delta\Gamma\big|_{\rm EOM} =  \Big(\underbrace{\dot{q}-\frac{q}{t}}_{{\cal O}(1/t)}+ \,{\cal O}(1/t^2)\Big)\delta q\Big|^{t_f} = {\cal O}(1/t)\underbrace{\delta q}_{\rm finite}\Big|^{t_f} = 0
\end{equation}
vanishes for all variations preserving the boundary conditions. 
The two problems mentioned in the previous Section indeed are resolved by the improved action \eqref{eq:dresden6} with \eqref{eq:dresden10}.

The considerations above apply in the same way to the Hamiltonian \eqref{eq:dresden1} with a more general class of half-binding potentials $V(q)$. In particular, the (manifestly positive) potential $V(q)$ is required to be monotonically decreasing, and to vanish faster than $1/q$ for large $q$. Going through the same steps as above is straightforward. Other generalizations, e.g.~to non-monotonic potentials or potentials with Coulomb-like behavior, may involve technical refinements, but the general procedure is always the same: one has to solve the Hamilton-Jacobi equation \eqref{eq:dresden8} to obtain the correct counterterm $\CT$ in \eqref{eq:dresden6}.

\section{Comparison with black hole partition functions}\label{se:3}

The same issues as in the previous Section arise when evaluating BH partition functions. Probably the simplest non-trivial model is 2-dimensional dilaton gravity (cf.~e.g.~\cite{Grumiller:2002nm} 
for recent reviews),
\begin{multline}\label{Action}
  I[g,X] = - \frac{1}{16\pi G_2}\,\int_{\MM} \!\! d^{\,2}x \,\sqrt{g}\, \left( X\,R - U(X)\,
		\left(\nabla X\right)^2 - 2 \, V(X) \raisebox{12pt}{~}\right) \\ 
		-  \frac{1}{8\pi G_2}\, \int_{\dM} \bns dx \, \sqrt{\gamma}\,X\,K ~.
\end{multline}
An explanation of the notation can be found in \cite{Grumiller:2007ju}. 
The boundary term in \eqref{Action} is the dilaton gravity analog of the Gibbons-Hawking-York boundary term. The latter arises in quantum mechanics if one converts the action $I=\int dt[-q\dot{p}-H(q,p)]$ into standard form, but it is {\em not} related to the Hamilton-Jacobi counterterm. 
It was shown first (second) in the second (first) order formulation \cite{Grumiller:2007ju} (\!\!\cite{Bergamin:2007sm}) that the improved action is given by
\begin{equation}\label{ActionConclusion}
  \Gamma[g,X] = I[g,X] + \frac{1}{8\pi G_2}\, \int_{\dM} \bns dx \,\sqrt{\gamma}  \,  \CT(X) \,,
\end{equation}
with the solution of the Hamilton-Jacobi equation ($V(X)\leq 0$)
\begin{equation}
\label{eq:C}
\CT(X) = \Big(-2 e^{-\int^X \!dy \,U(y)}\int^X \!\!dy\,V(y)\,e^{\int^y\!dz\,U(z)}\Big)^{1/2}\,.
\end{equation}
The lower integration constant in the integrals over the function $U$ is always the same and therefore cancels; the lower integration constant in the remaining integral represents the ambiguity mentioned in footnote \ref{fn:1}. 
\footnote{Solving \eqref{eq:C} for $V$ yields $V=-1/2 \,[(\CT^2)^\prime+\CT^2 U]$, which reveals that the Hamilton-Jacobi counterterm $\CT(X)$ is the supergravity pre-potential (up to a numerical factor) \cite{Grumiller:2002nm}. A similar (pseudo-)su\-per\-symmetric story exists in quantum mechanics \cite{Townsend:2007aw}. Cf.~also \cite{deBoer:1999xf}.}

The BH partition function based upon the improved action \eqref{ActionConclusion},
\begin{equation}\label{PartitionFunction2}
  \ZZ  = \int \CurlD g \,\CurlD X \, \exp\Big(- \frac{1}{\hbar}\,\Gamma[\,g,X]\Big) \approx \exp\Big(- \frac{1}{\hbar}\,\Gamma[\,g_{\rm cl},X_{\rm cl}]\Big)\,,
\end{equation}
by standard methods establishes the BH free energy. The asymptotic part of the counterterm \eqref{eq:C} leads to the correct asymptotic charges for BHs with (essentially) arbitrary asymptotic behavior, and to consistency with the first law of thermodynamics (which is non-trivial \cite{Davis:2004xiNote}). The finite part of the counterterm \eqref{eq:C} allows a quasi-local description of BH thermodynamics \cite{Grumiller:2007ju}. 

Perhaps one might exploit the formal analogy between BH partition functions and quantum mechanical partition functions described in this work to construct interesting condensed matter analogs \cite{Novello:2002qg} 
mimicking thermodynamical aspects of BHs.

\section*{Acknowledgments}

I am grateful to Roman Jackiw and Robert McNees for enjoyable discussions. I thank Luzi Bergamin and Ren{\'e} Meyer for reading the manuscript and Wolfhard Janke for valuable administrative help.

This work is supported in part by funds provided by the U.S. Department of Energy (DoE) under the cooperative research agreement DEFG02-05ER41360 and by the project MC-OIF 021421 of the European Commission under the Sixth EU Framework Programme for Research and Technological Development (FP6). Part of my travel expenses where covered by funds provided by the conference ``Path integrals -- New Trends and Perspectives'' at the Max-Planck Institute PKS in Dresden.


\providecommand{\href}[2]{#2}\begingroup\raggedright\endgroup

\end{document}